\def\b{\bibitem}
\def\boldphi{\mbox{\boldmath $\phi$}}
\def\boldvarphi{\mbox{\boldmath $\varphi$}}
\documentstyle[aps,prl,floats,epsf]{revtex}
\begin{document}
\def\SNG{{\em Physical Review Style and Notation Guide}}
\def\LUG {{\em \LaTeX{} User's Guide \& Reference Manual}}
\def\btt#1{{\tt$\backslash$\string#1}}%
\def\REVTeX{REV\TeX}
\def\AmS{{\protect\the\textfont2
        A\kern-.1667em\lower.5ex\hbox{M}\kern-.125emS}}
\def\AmSLaTeX{\AmS-\LaTeX}
\def\BibTeX{\rm B{\sc ib}\TeX}
\twocolumn[\hsize\textwidth\columnwidth\hsize\csname@twocolumnfalse%
\endcsname
\title{Influence of rare regions on magnetic quantum phase transitions}
\author{Rajesh Narayanan$^1$, Thomas Vojta$^{1,2}$, D. Belitz$^1$,
                                                and T.R. Kirkpatrick$^{3}$}
\address{$^1$Department of Physics and Materials Science Institute,
                                      University of Oregon, Eugene, OR 97403}
\address{$^2$Institut f{\"u}r Physik, TU Chemnitz, D-09107 Chemnitz, FRG}
\address{$^3$Institute for Physical Science and Technology, and Department
             of Physics, University of Maryland, College Park, MD 20742}
\date{\today}
\maketitle
\begin{abstract}
The effects of quenched disorder on the critical properties of itinerant 
quantum magnets are considered. Particular attention is paid to locally 
ordered rare regions that are formed in the presence of quenched 
disorder even when the bulk system is still in the nonmagnetic phase. 
It is shown that these local moments or instantons destroy the previously 
found critical fixed point in the case of antiferromagnets. In the case 
of itinerant ferromagnets, the critical behavior is unaffected by the 
rare regions due to an effective long-range interaction between the order 
parameter fluctuations.
%
%
\end{abstract}
]
Rare regions and their influence on observables is an important,
if intricate, aspect of systems with quenched disorder. An
effect that has been known for a long time is the formation of a Griffiths
region\cite{Griffiths}. To explain this, let us consider a 
ferromagnet for definiteness. Disorder will decrease the critical 
temperature from its clean value, $T_c^0$, 
to a value $T_c < T_c^0$ in the disordered system. In the temperature region
$T_c < T < T_c^0$ the system does not display global order, but
one will find regions that are devoid of any impurities and hence 
show local magnetic order. The probability of 
finding such a `rare region' in general decreases exponentially with its size.
The resulting magnetization fluctuations have very slow dynamics. They 
are often called `local moments' or `instantons', and they lead to a 
nonanalytic free energy 
for all temperatures below $T_c^0$, even though no long-range order develops
until the temperature reaches $T_c$. For generic classical systems this is a
weak effect, the singularity being only an essential one. An important
exception is the model studied by McCoy and Wu\cite{McCoyWu}, which is a
two-dimensional ($2$-$d$) Ising model with random bonds in one direction, but 
identical bonds in the second direction. The infinite correlation 
of the disorder in this model leads to much stronger effects, with 
the average magnetic susceptibility diverging in a finite-width temperature 
region above $T_c$. The transition at $T_c$ is nevertheless sharp. The
divergence of the average susceptibility for $T>T_c$ is caused by atypical
fluctuations in the susceptibility distribution, and the averaged order
parameter becomes nonzero only for $T<T_c$. The temperature 
region $T_c < T < T_c^0$ is known as
a Griffiths region. Little is known about the influence of rare regions
on the critical behavior at $T_c$, and in the conventional theory of the
critical behavior of disordered magnets\cite{Grinstein} the rare regions 
are neglected.

Recent work\cite{Dotsenko} on a random-$T_c$ classical Ising model has 
suggested that the effects of the rare regions go beyond the formation 
of a Griffiths region, even in this simple model where the conventional
theory\cite{Grinstein} predicts standard power-law critical behavior. 
These authors showed that the conventional theory is unstable with 
respect to perturbations that break the replica symmetry. By approximately 
taking into account the rare regions, they found a new term in the
action that actually induces such perturbations. In some systems 
replica symmetry breaking is believed to be associated with activated,
i.e. non-power law, critical behavior. Although no final conclusion about 
the fate of the transition could be reached, Ref.\ \onlinecite{Dotsenko}
thus raised the possibility that the random-$T_c$ classical
Ising model shows activated critical behavior as a result of rare-region
effects, as is believed to be the case for the random-field classical
Ising model\cite{random-field}. 

The problem of rare regions is even less well investigated for the case
of quantum phase transitions, i.e. transitions that occur at $T=0$
as a function of some non-thermal control parameter\cite{QPT}.
An important exception to this are certain $1$-$d$ systems. 
Fisher\cite{DSF} has investigated quantum Ising spin chains in a transverse 
random field, which is closely related to the classical $2$-$d$ McCoy-Wu
model (with time playing the role of the dimension along which the
disorder is correlated). He found activated critical
behavior due to rare regions, which has been confirmed by numerical
simulations\cite{APY1d}. Recent simulations\cite{APY2d} suggest that this
type of behavior may not be restricted to $1$-$d$ systems, raising
the possibility that exotic critical behavior dominated by rare regions
may be generic in quenched disordered quantum systems.

Apart from their relevance for disordered magnets and their 
critical properties,
rare regions are believed to be a crucial ingredient for understanding other
systems with quenched disorder. For instance, it has been proposed that
a complete understanding of the properties of doped semiconductors, and of the
metal-insulator transitions that are observed in such systems as a function
of doping, requires the consideration of 
local moments\cite{BhattLee,BhattFisher,Milovanovicetal}.

In this Letter we study this important problem analytically for quantum phase
transitions in $d>1$. We concentrate on magnetic transitions, 
and contrast the cases of itinerant ferromagnets (FMs) and 
antiferromagnets (AFMs), respectively. We find that for the
latter, the rare regions destroy the critical fixed point (FP) found in
a previous study\cite{afm}, and thus have a 
profound effect on the critical behavior. In contrast, for itinerant
FMs we find that, for certain realizations of the disorder, 
the previously found critical behavior\cite{fm} is
stable with respect to rare regions, due to an effective long-range
interaction between the spin fluctuations. In addition, we find that
the ultimate effects of the rare regions depend  on how the
disorder is realized in a particular system. Therefore, no generally valid 
conclusions are possible, and the effects of rare regions
must be studied carefully and separately for each system
under consideration.

Let us first consider the case of itinerant quantum antiferromagnets. Our
starting point is the same
as in Ref.\ \onlinecite{afm}, namely Hertz's action, which is
a $\phi^4$-theory for a $p$-component order parameter field $\boldphi$
whose expectation value is proportional to the staggered magnetization.
The action reads
\begin{mathletters}
\label{eqs:1}
\begin{equation}
S = \int dx\,dy\ {\boldphi}(x)\,\Gamma(x,y)\,{\boldphi}(y)
    + u\int dx\ \left({\boldphi}(x)\cdot{\boldphi}(x)\right)^2\,.
\label{eq:1a}
\end{equation}
Here $x\equiv ({\bf x},\tau)$ comprises position ${\bf x}$ and imaginary
time $\tau$, and $\int dx \equiv \int d{\bf x}\int_{0}^{1/T}d\tau$. We
use units such that $\hbar = k_{\rm B} = 1$. $\Gamma(x,y)$ is the bare 
two-point vertex function, whose Fourier transform is
\begin{equation}
\Gamma({\bf q},\omega_n) = (t + {\bf q}^2 + \vert\omega_n\vert)/2\quad.
\label{eq:1b}
\end{equation}
\end{mathletters}%
Here $t$ denotes the distance from the critical point, ${\bf q}$ is the
wavevector, $\omega_n$ is a bosonic Matsubara frequency, and we measure
both ${\bf q}$ and $\omega_n$ in suitable units. 

Disorder is introduced by making $t$ a random function of position,
$t = t_0 + \delta t({\bf x})$, where $\delta t({\bf x})$ obeys a 
Gaussian distribution with zero mean and variance $\Delta$. The standard 
procedure is to integrate out the `random mass' $\delta t({\bf x})$ 
by means of the replica trick\cite{Grinstein}, which produces a term of 
order ${\boldphi}^4$ with coupling constant $\Delta$, in addition
to the ordinary quantum fluctuation term in Eq.\ (\ref{eq:1a}) with coupling 
constant $u$. The resulting theory does not easily allow for saddle-point
solutions that are inhomogeneous in space, and to incorporate rare regions
into it would be very difficult. We therefore follow a different procedure.
In analogy to Ref.\ \onlinecite{Dotsenko}, we consider inhomogeneous
saddle-point solutions of the theory for a {\em fixed} realization of the
disorder. The inhomogeneity comes about since $\delta t({\bf x})$ has
`troughs' that make $t<0$ in some region in space, even though $t_0>0$. 
Troughs that are
sufficiently deep and wide support locally nonzero saddle-point solutions.
These regions we will refer to as `islands'. Outside of the islands,
the solution is exponentially small. This means that for a system with
$N$ islands, and in the case of an Ising model ($p=1$), there will be $2^N$ 
almost degenerate saddle-point solutions that can be constructed by 
considering all possible distributions of the sign of the order parameter
on the islands. For $p>1$ there is a whole manifold of almost degenerate
saddle points.

Let ${\bf\Phi}({\bf x})$ be one of these saddle-point solutions, and let
us consider fluctuations about it, 
$\boldphi(x) = {\bf\Phi}({\bf x}) + \boldvarphi(x)$\cite{DynamicsFootnote}.
The different saddle points are
far apart in configuration space and separated by large energy barriers.
If we restrict ourselves to small fluctuations about each saddle point, 
we can therefore write
the partition function approximately as the sum of all contributions
obtained from the vicinity of each saddle point,
\begin{equation}
Z \approx \int D[{\bf\Phi}({\bf x})]\,P[{\bf\Phi}({\bf x})] 
   \int_{<} D[\boldvarphi(x)]\,
   e^{-S[{\bf\Phi}({\bf x}) + \boldvarphi(x)]},
\label{eq:2}
\end{equation}
where $P[{\bf\Phi}({\bf x})]$ denotes the distribution of saddle points, and 
$\int_<$ indicates an integration over small fluctuations only.
It is clear that this approximation takes into account effects that
one would call `non-perturbative' in a standard treatment of quenched
disorder. Also, consistent with our approximations, it can be shown that
the inhomogeneous saddle-point solutions lead to a lower free energy than
the homogeneous saddle point $\boldphi(x) \equiv 0$.

Performing the integration over the $\bf\Phi$ in Eq.\ (\ref{eq:2}) explicitly 
is very difficult, and the result will in general depend on the properties
of the distribution function $P$, which in turn depend on the details of
the miscroscopic disorder. However, a very basic observation simplifies
our task: The ${\bf\Phi}({\bf x})$ represent static randomness,
and the average over this randomness is performed for the partition function.
That is, we are dealing with static, annealed disorder. This is physically 
sensible, as the local moments are a self-generated part of the system and 
therefore in equilibrium with the rest of the degrees of 
freedom\cite{DynamicsFootnote}. In addition, of course,
there is quenched disorder due to the underlying random mass term. This we
handle by means of the replica trick. If we assume
that the distribution of the $\bf\Phi$ is short-range correlated
(which will be the case for certain classes of realizations of the disorder,
but not for others), we can immediately write down the effective action up 
to and including terms of $O(\boldvarphi^4)$:
\begin{eqnarray}
S_{\rm eff}&=&\sum_{\alpha} \int dx\,dy\ \boldvarphi^{\alpha}(x)\,
                   \Gamma_{0}(x,y)\,\boldvarphi^{\alpha}(y)
\nonumber\\
&&+ u\sum_{\alpha}\int dx\ \left(\boldvarphi^{\alpha}(x)\cdot
                                    \boldvarphi^{\alpha}(x)\right)^2
\nonumber\\
&&- \sum_{\alpha,\beta}(\Delta + w\,\delta_{\alpha\beta})\int dx\,dy\,
  \delta({\bf x}-{\bf y})\,
  \left(\boldvarphi^{\alpha}(x)\right)^2\,
\nonumber\\
&&\qquad\qquad\qquad\qquad\times \left(\boldvarphi^{\beta}(y)\right)^2
 + O(\boldvarphi^6)\quad.
\label{eq:3}
\end{eqnarray}
Here $\Gamma_0$ is the Gaussian vertex, Eq.\ (\ref{eq:1b}), with $t=t_0$,
$\Delta$ is the variance of the Gaussian random mass distribution,
and $\alpha$ and $\beta$ are replica indices. $w$ is the coupling constant
of the annealed disorder term.
We have also derived Eq.\ (\ref{eq:3}) by means of a detailed technical
procedure which will be reported elsewhere\cite{ustbp}. The technical
derivation shows that $w$ has the form $w = u^2\,v$, with $v$ characteristic
of the distribution $P$, and it also yields terms of $O(\boldvarphi^6)$ and
higher. These turn out to be less relevant for the critical behavior than
the quartic terms shown in Eq.\ (\ref{eq:3}).
Notice that the annealed disorder contribution becomes indistinguishable
from the usual $\boldvarphi^4$ or $u$-term in the case of a classical
transition. This is the reason why the authors of Ref.\ 
\onlinecite{Dotsenko}, who studied classical magnets, considered replica
symmetry breaking in order to describe nontrivial effects of the rare regions.
In the quantum case we get a nontrivial effect even at the level of a
replica symmetric theory, which means that the influence of rare regions
on quantum transitions is stronger.

To discuss the properties of the effective action, Eq.\ (\ref{eq:3}), we
proceed as in Ref.\ \onlinecite{afm}. We consider $d = 4 - \epsilon$ space
dimensions and $\epsilon_{\tau}$ time dimensions, and control perturbation
theory by means of a double expansion in $\epsilon$ and 
$\epsilon_{\tau}$\cite{CardyBoyanovsky}. Defining 
${\bar w} = w\,T^{-\epsilon_{\tau}}$, and putting $T=0$, we obtain the 
following renormalization group (RG) flow equations to one-loop order,
\begin{mathletters}
\label{eqs:4}
\begin{eqnarray}
\frac{du}{dl} = (\epsilon - 2\epsilon_{\tau})u - 4(p+8)u^2 + 48u\Delta\quad,
\label{eq:4a}\\
\frac{d\Delta}{dl} = \epsilon\Delta + 32\Delta^2 - 8(p+2)u\Delta 
                                    + 8p\Delta{\bar w}\quad,
\label{eq:4b}\\
\frac{d{\bar w}}{dl} = (\epsilon - 2\epsilon_{\tau}){\bar w} + 4p{\bar w}^2
                       - 8(p+2)u{\bar w} + 48\Delta{\bar w}\quad.
\label{eq:4c}
\end{eqnarray}
\end{mathletters}%
An analysis of Eqs.\ (\ref{eqs:4}) shows that they possess eight FPs. Four
of them have a vanishing FP value of ${\bar w}$, ${\bar w}^* = 0$, and have
been discussed before in Ref.\ \onlinecite{afm}. Of particular interest is
the nontrivial critical FP $u^* = (\epsilon + \epsilon_{\tau})/16(p-1)$,
$\Delta^* = [(4-p)\epsilon + 4(p+2)\epsilon_{\tau}]/64(p-1)$, ${\bar w}^* = 0$,
which on the ${\bar w}=0$ hypersurface is stable for $p$ smaller than some
$p_c$. To one-loop order, and for $\epsilon = \epsilon_{\tau}$, $p_c = 16$. 
A linear stability analysis reveals that the third eigenvalue,
$\lambda_{\bar w} = (4-p)(\epsilon + 4\epsilon_{\tau})/4(p-1)$, is positive for
$p<4$. In the most interesting case $p=3$, ${\bar w}$ is thus a relevant 
operator with respect to this FP, which means that the rare regions destroy 
the FP. It is, however, interesting to note that for $p>4$ the FP is stable
and describes power-law critical behavior. 
There also are four FPs with
${\bar w}^*\neq 0$. They are all unstable except for one with
${\bar w}^* = (p-4)(\epsilon + 4\epsilon_{\tau})/8p(10-p)$, which is negative
for $p<4$. Since the bare value of ${\bar w}$ is positive, and the structure
of the flow equations does not allow for ${\bar w}$ to change sign, this FP
is unphysical. There is thus no new FP for $p<4$, and a numerical solution 
of the flow equations reveals runaway flow in all of physical parameter space.

We conclude that for $p<4$ the AFM long-range order found in 
Ref.\ \onlinecite{afm} is unstable against effects induced by rare regions, 
a result that is consistent
with the previous suggestion that AFM long-range order is strongly suppressed
by quenched disorder\cite{BhattLee}. However, other possibilities exist. For 
instance, there could be a transition to a long-range ordered state,
but with activated critical behavior which manifests itself as runaway flow in
a perturbative RG calculation. The viability of this latter suggestion is
underscored by the fact that a calculation of the local moment contribution
to the order parameter susceptibility yields 
$\chi_{\rm LM}(T) \sim 1/T$ \cite{ustbp}. This is 
similar to Fisher's $1$-$d$ result $\chi (T) \sim 1/T^{\gamma}$ 
with $\gamma < 1$ \cite{DSF}. (Our exponent value of unity is 
a result of our saddle-point approximation for the local moments.) This shows
that we are really describing a Griffiths region, which was shown in
Ref.\ \onlinecite{DSF} to lead to a transition with activated critical
behavior in $d=1$. A third possibility is that a conventional critical FP
exists, but cannot be described with perturbative RG
methods. This possibility is consistent with the stability of conventional 
critical behavior against ${\bar w}$ for $p>4$, as discussed
above.

We now turn to the case of itinerant ferromagnets, which constitute an
interesting contrast to the AFM case. In Ref.\ \onlinecite{fm}
it was shown that a description of itinerant FMs that neglects rare
regions leads to an action that has the same form as Eq.\ (\ref{eq:3})
with $w=0$, except that the bare two-point vertex function reads
\begin{equation}
\Gamma_0^{\rm FM}({\bf q},\omega_n) = \left(t_0 + \vert{\bf q}\vert^{d-2} 
   + {\bf q}^2 + \vert\omega_n\vert/{\bf q}^2\right)/2\quad,
\label{eq:5}
\end{equation}
and that the field $\boldvarphi(x)$ now describes ferromagnetic fluctuations.
There are two crucial, and related, differences between Eq.\ (\ref{eq:5})
and its AFM counterpart. The first one is the structure of the frequency
dependence, which enters as $\vert\omega_n\vert/{\bf q}^2$\cite{Hertz}
and reflects the diffusive nature of the spins in a disordered environment.
In Ref.\ \onlinecite{fm} it was shown that the same diffusive spin dynamics
leads to the $\vert{\bf q}\vert^{d-2}$ term, which dominates the usual
gradient squared term as long as $d<4$. In the original treatment of quantum
FMs by Hertz\cite{Hertz}, loop corrections
would have been required to find this term, while the method
of Ref.\ \onlinecite{fm} builds it into the bare theory. Consequently, the 
correlations between the spin density fluctuations are effectively long-ranged,
a feature that is well known to stabilize the Gaussian critical 
behavior\cite{FisherMaNickel}. Indeed, it was shown in Ref.\ \onlinecite{fm}
that the Gaussian critical behavior, with $\eta = 4-d$, $\nu = 1/(d-2)$, 
$\gamma = 1$, and $z = d$, is stable for $2<d<4$. Here $\eta$, $\nu$, and
$\gamma$ are the usual critical exponents, and $z$ is the dynamical critical
exponent. They can all be simply read off Eq.\ (\ref{eq:5}). The exponents
$\beta$ and $\delta$ were also determined in Ref.\ \onlinecite{fm}, their
values in $d=3$ are $\beta=2$, $\delta=3/2$. The remarkable claim of
Ref.\ \onlinecite{fm} was that these exponent values, which in $d=3$ 
are very different from both mean-field values and classical Heisenberg
values, constitute the {\em exact} critical behavior of itinerant quantum
FMs.

An obvious question is whether this claim survives the consideration of
rare regions. To answer this, we perform an analysis analogous to the one
for AFMs above. A simple way to incorporate the rare regions into the
action is to write the quenched disorder or $\Delta$-term in the action
as a random mass in the Gaussian vertex (i.e., to `undo' the integrating-out 
of the random mass), to construct inhomogeneous saddle-point solutions
and expand about them, and then to integrate over the manifold of saddle
points as in the AFM case. Clearly, this leads to a $w$-term in the action,
like in Eq.\ (\ref{eq:3}). We have derived the same result starting
from a more microscopic formulation. We will report the details
of the derivation elsewhere\cite{ustbp}, here we mention only one important
point: After Eq.\ (\ref{eq:3}) we mentioned that $w$ is proportional
to $u^2$. Since $u$ in the FM case is wavenumber dependent and diverges in
the short-wavelength limit (i.e., its bare scale dimension is 
negative)\cite{fm}, this raises the question whether the bare value of $w$
is finite. The answer is affirmative, since the $w$-term arises from field
configurations that are nonzero only on islands. The $u$ that contributes to
the bare value of $w$ therefore has to be taken at wavenumbers that are on
the order of a typical inverse island size, and hence is finite. Once again
it is important here that the island size distribution falls off exponentially
for large sizes. We can thus treat $w$ as a number.

Now let us perform a power counting analysis to determine the stability
properties of the Gaussian FP. Assigning a length $L$ a scale dimension
$[L] = -1$, the scale dimension of the imaginary time is $[\tau] = -z = -d$.
For the scale dimension of the field we find $[\boldvarphi(x)] = (d+2)/2$.
The scale dimensions of both $w$ and $\Delta$ then become
$[w] = [\Delta] = 4-d$, i.e. they are irrelevant with respect to the
Gaussian FP for $d<4$, and marginal in $d=4$. Terms of higher than quartic
order in $\boldvarphi$ that are produced by a technical derivation of the
effective action\cite{ustbp} turn out to also be irrelevant.
We thus conclude that the
FM critical behavior determined previously\cite{fm} is {\em stable} against
rare regions physics, in sharp contrast to the AFM case. The reason for
this qualitative difference is the effective long-ranged interaction between
the order parameter fluctuations (as expressed in Eq.\ (\ref{eq:5}) and
in the value of the exponent $\eta$), which is sufficient to suppress all
disorder fluctuations, including the ones due to rare regions. By the same
arguments, the FM Gaussian FP is also stable against replica symmetry breaking.

We conclude with one additional remark. One might ask why the rare regions 
or local moments don't cut off the singular wavenumber dependences 
$\vert{\bf q}\vert^{d-2}$ and $\vert\omega_n\vert/{\bf q}^2$
in the Gaussian vertex, Eq.\ (\ref{eq:5}). The reason why this does not happen
is that both singularities are consequences of spin diffusion, which in
turn is a consequence of the spin conservation law. The rare regions
ultimately derive from a spin-independent disorder potential, which
clearly cannot destroy spin conservation. We note, however, that the
above arguments are restricted to a tree-level analysis of our effective
field theory. Although the effective theory is a sophisticated one, which
at tree level contains many effects that would require loops in more
standard treatments, we of course cannot exclude the possibility that 
loop corrections might lead to qualitatively new terms in the action.
If such new terms included a RG-generated spin dependent potential,
then this might change our conclusions. However, at such a level of
analysis one would also have to include effects due to interactions
between the rare regions, which we have mostly neglected. Such interactions
are known to weaken the effects of the rare region\cite{BhattFisher},
but in general it is not known by how much. 


We gratefully acknowledge helpful discussions with Ferdinand Evers and John 
Toner. This work was supported by the NSF under grant Nos. DMR-98-70597 and
DMR--96--32978, and by the DFG under grant No. SFB 393/C2.

\end{document}